\documentclass[aps,prl,twocolumn,groupedaddress]{revtex4-1}
\usepackage{calrsfs}
\usepackage{mathrsfs}
\usepackage{graphicx}% Include figure files
\usepackage{dcolumn}% Align table columns on decimal point
\usepackage{bm}% bold math
\usepackage{hyperref}
\usepackage{xcolor}
\hypersetup{colorlinks=true,linkcolor=red}
\usepackage{color}
\usepackage{fancyhdr}
\usepackage{url}
\usepackage{amssymb,amsfonts,amsmath,amsbsy,bm,t1enc,lipsum,latexsym}
\usepackage{changes}

\newcommand{\eqt}[1]{``#1''}

\begin{document}

\normalem

\preprint{APS/123-QED TODO}

\title{Higher order mode suppression in high-\emph{Q} anomalous dispersion SiN microresonators for temporal dissipative Kerr soliton formation}

\author{A. Kordts}
\affiliation{{\'E}cole Polytechnique F{\'e}d{\'e}rale de Lausanne (EPFL), CH-1015 Lausanne, Switzerland}

\author{M. H. P. Pfeiffer}
\affiliation{{\'E}cole Polytechnique F{\'e}d{\'e}rale de Lausanne (EPFL), CH-1015 Lausanne, Switzerland}

\author{H. Guo}
\affiliation{{\'E}cole Polytechnique F{\'e}d{\'e}rale de Lausanne (EPFL), CH-1015 Lausanne, Switzerland}

\author{V. Brasch}
\affiliation{{\'E}cole Polytechnique F{\'e}d{\'e}rale de Lausanne (EPFL), CH-1015 Lausanne, Switzerland}

\author{T. J. Kippenberg}
\email[E-mail: ]{tobias.kippenberg@epfl.ch}
\affiliation{{\'E}cole Polytechnique F{\'e}d{\'e}rale de Lausanne (EPFL), CH-1015 Lausanne, Switzerland}

\begin{abstract}
High-\emph{Q} silicon nitride (SiN) microresonators enable optical Kerr frequency comb generation on a photonic chip and have recently been shown to support fully coherent combs based on temporal dissipative Kerr soliton formation. For bright soliton formation it is necessary to operate SiN waveguides in the multimode regime so as to produce anomalous group velocity dispersion. This can lead to local disturbances of the dispersion due to avoided crossings caused by coupling between different mode families, and therefore prevent the soliton formation. Here we demonstrate that a single mode \eqt{filtering} section inside high-\emph{Q} resonators enables to efficiently suppress avoided crossings, while preserving high quality factors (\emph{Q} ${\sim 10^6}$). We demonstrate the approach by single soliton formation in SiN resonators with filtering section.
\end{abstract}

\pacs{42.65.Ky, 42.65.Tg, 42.60.Da, TODO}

\maketitle

\section{Introduction}

Silicon nitride (SiN) integrated waveguides are an ideal platform for on-chip nonlinear optics \cite{Moss2013Cmos,Levy2010Cmos}, which advance diverse research topics such as supercontinuum generation \cite{Zhao15octaveContinium,Epping15continium,Halir12continuum} and microresonator Kerr frequency comb generation \cite{Kippenberg2011combs}. The latter represents a technology that enables an optical frequency comb with the mode spacing in the microwave range and with large bandwidth reaching one octave \cite{Haye2011octaveComb,Okawachi2011octave}.
Applications of low phase noise comb states in SiN microresonators \cite{Herr2012UniversalFormation} so far include coherent communication \cite{Pfeiffle2014terabitCom} as well as arbitrary waveform generation \cite{Weiner2011LineByLine}.
Recently the demonstration of temporal dissipative Kerr soliton formation in microresonators, in crystalline resonators \cite{Herr2013firstSoliton,Liang2015highPurityComb}, photonic chip-based SiN microresonators \cite{Brasch2014Cherenkov} and in monolithic silica micro-disks \cite{Yi2015soliton}, further provides a reliable and novel method for the generation of fully coherent and broadband frequency combs with smooth spectral envelope, and the generation of ultrafast and ultrashort femtosecond pulses.
This enables novel applications such as low-noise microwave generation \cite{Liang2015highPurityComb,Yi2015soliton,Jost2015cycles}, coherent data transmission \cite{Pfeiffle2014terabitCom,Pfeifle15solitonCom} and ultrafast spectroscopy.
In addition, it has been shown that the spectral bandwidth can be substantially increased into the normal dispersion regime \cite{Brasch2014Cherenkov} using soliton induced Cherenkov radiation\cite{Akhmediev1995Cherenkov,Coen2013octaveLLE}.

However it was observed that locally altered dispersion can prevent the soliton formation through the interaction between different mode families supported by the resonator; the problem was first addressed in \cite{Herr2013avoidedCrossings}. For certain frequencies two modes belonging to different families can be almost resonant and thus a minute coupling between both, e.g. through waveguide imperfections, can result in the formation of hybrid modes with shifted resonance frequencies. This results in a local defect in the resonator dispersion, which is termed \eqt{avoided modal crossing}.

\begin{figure}[tb]
\centering
\includegraphics[width=\linewidth]{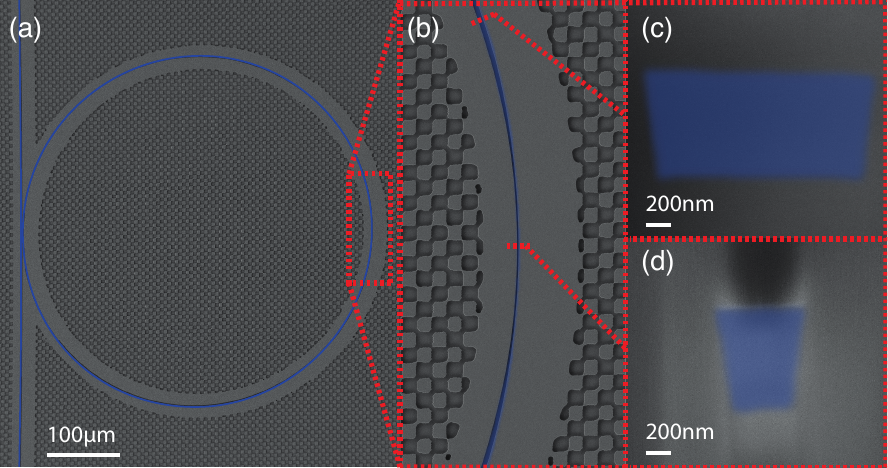}
\caption{Scanning electron microscopy (SEM) images of a dispersion optimized ${\rm 240}\text{ }\mu {\rm m}$ SiN microring resonator: (a) picture of the whole resonator; (b) the mode filtering section with an adiabatic tapered section (length ${\rm 130}\text{ } \mu {\rm m}$). (c, d) cross sections of the ring resonator's waveguide (dimensions $0.8 \times 1.65\text{ }\mu {\rm m}$), with (d) the taper waist (dimensions $0.8 \times 0.6\text{ }\mu {\rm m}$).
}
\label{fig:DesignAndFab}
\end{figure}

While detrimental for dissipative Kerr soliton formation, such an avoided modal crossing can provide local anomalous group dispersion (GVD) in microresonators such that it also initiates generation of Kerr frequency combs in an otherwise normal GVD regime\cite{Savchenkov2012OvermodedResonators}. Dual-ring geometries were also used to induce controllable avoided modal crossing between microresonators (in normal GVD regime) and to generate frequency combs with adjustable free spectral ranges (FSRs) \cite{2014modecoupling,miller2015dualring}.

In this letter, we present a novel, yet simple method to suppress higher order mode families in SiN microresonators by introducing a mode filtering section into the ring microresonator.
We show that by inserting an adiabatic transition to a single mode waveguide inside the resonator, avoided crossings in the resonator can be strongly reduced, while preserving the anomalous GVD as well as the high quality factor $(\emph{Q} \sim 10^6)$ of the fundamental modes.
We further demonstrate, with the novel resonator design, the generation of broadband frequency combs based on single dissipative Kerr soliton formation.

\section{Device design and fabrication}

\begin{figure}[t]
\centering
\includegraphics[width=\linewidth]{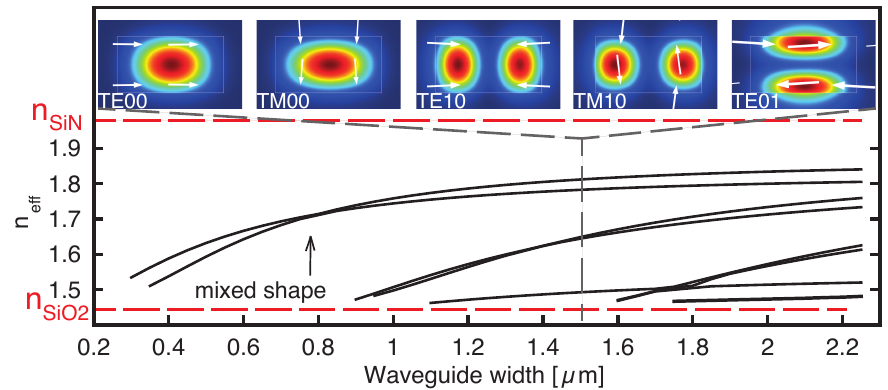}
\caption{Effective refractive index as a function of the waveguide width for the full set of mode families, at the free space wavelength $\lambda_{0} = $ 1.55 ${\mu {\rm m}}$; the mode profiles including the electric field vectors are shown as insets for the width of ${1.5 ~\mu {\rm m}}$; the waveguide height is 0.8 ${\mu {\rm m}}$ and the resonator's FSR is 100 GHz. An example region of mixed-shape mode profiles is marked.
}
\label{fig:SimData}
\end{figure}

The novel design of the mode filtering section in SiN microresonators is realized by tapering down the multimode ring waveguide to single mode, see Fig. \ref{fig:DesignAndFab}.
The design is a trade-off between two criteria, namely on one hand, the filtering
section has to be sufficiently long and the waist sufficiently small such that higher
order mode families are cut-off; on the other hand, the taper length has to be sufficiently short, to preserve an overall resonator dispersion that yields still anomalous GVD in order to allow for dissipative Kerr soliton formation and frequency comb generation via four-wave-mixing processes \cite{Chembo2010anomalousDisp}.
The boundary of the tapered waveguide region is defined as:
\begin{equation}
w(\phi)=\frac{w_{\rm n}+w_{\rm t}}{2}-\frac{w_{\rm n}-w_{\rm t}}{2}\cos\left(\pi\left(\frac{2r\phi}{l_{t}}\right)^{\mathrm{3}}\right)
\end{equation}
where $l_{\rm t}$ is the taper length, $r$ is the resonator radius, $w_{\rm n}$ is the nominal width of the multimode waveguide, ${w_{\rm t}}$ is the minimal width at the taper waist.
${r \cdot \phi \in [-\frac{l_{\rm t}}{2},\frac{l_{\rm t}}{2}]}$ where $\phi$ is the angular coordinate and ${\phi = 0}$ indicating the position of the taper waist.

In order to reveal the effects of the filtering section, we simulated eigen-modes \cite{Oxborrow2007}, including the mode profile and the cavity resonance frequency as a function of the waveguide width for the full set of mode families, see Fig. \ref{fig:SimData}.
%Figure \ref{fig:SimData} shows the effective refractive index profile as a function of the waveguide width for the full set of mode families.
When the waveguide width is narrowed down below 0.8 ${\mu{\rm m}}$, higher order modes are cut-off, leaving only the two fundamental mode families (${\rm TE_{00}, TM_{00}}$).
Regions of \eqt{mixed-shape} mode profiles were also observed, which were carefully studied in \cite{Carmon2008}. Consequences are not further regarded in the present work.
The set of SiN microresonators studied in this work has a waveguide height of 0.8 ${\mu {\rm m}}$ and a nominal width of 1.65 ${\mu {\rm m}}$ such that anomalous GVD is produced over a wide wavelength span. The FSR is 100 GHz. The taper waist in the filtering section ranges from 0.45 ${\mu {\rm m}}$ to the nominal width and the taper length is fixed to be 130 ${\mu {\rm m}}$. It should be noted that the parameter of the waveguide width in this paper has an undetermined offset in the range of $\pm$30 nm, induced during the microfabrication process.

The tapered SiN microresonator devices were fabricated using the \emph{Photonic
Damascene Process} \cite{Pfeiffer2015Damascene}. The waveguide pattern
was defined using electron beam lithography and transferred by dry
etching into the silicon dioxide substrate. Additionally a dense checkerboard
pattern was structured around the waveguide by photolithography
and dry etching to release the stress in the deposited SiN thin film.
After the deposition of the SiN film the excess material was removed
using chemical mechanical polishing. In the last steps the waveguide was clad
with oxide, annealed and separated into chips.

\section{Experiments and results}

\begin{figure}[t]
\includegraphics[width=\linewidth]{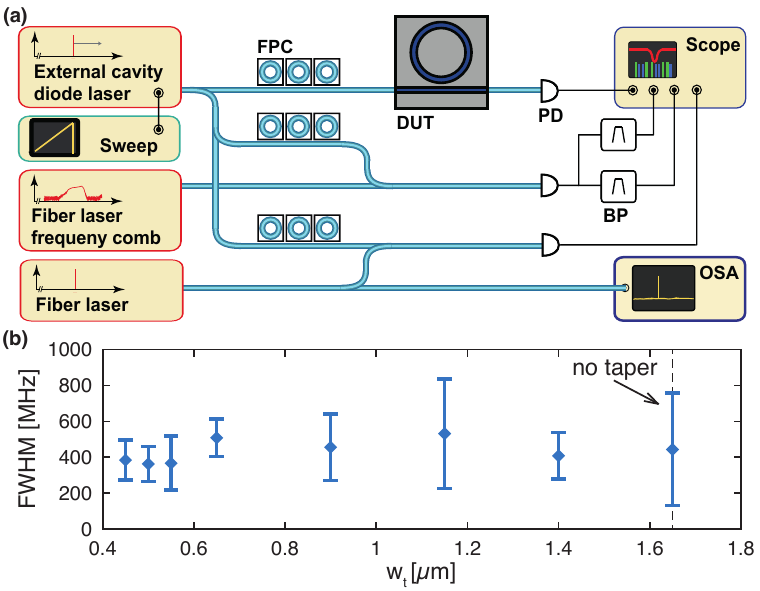}
\caption{(a) Setup of frequency comb assisted tunable laser spectroscopy\citep{Riemensberger2012dispMeasurement}; DUT: device under test, ECDL: external cavity diode laser, PD: photodiode, BP: band pass filter, OSA: optical spectral analyzer; (b) mean value and standard deviation of the resonance linewidth, over the measurement wavelength range 1.51 -- 1.61 ${\mu {\rm  m}}$, as a function of the width at the tapered waist $w_t$.}
\label{fig:SetupAndQs}
\end{figure}

\begin{figure*}[t]
\includegraphics[clip,scale=0.95]{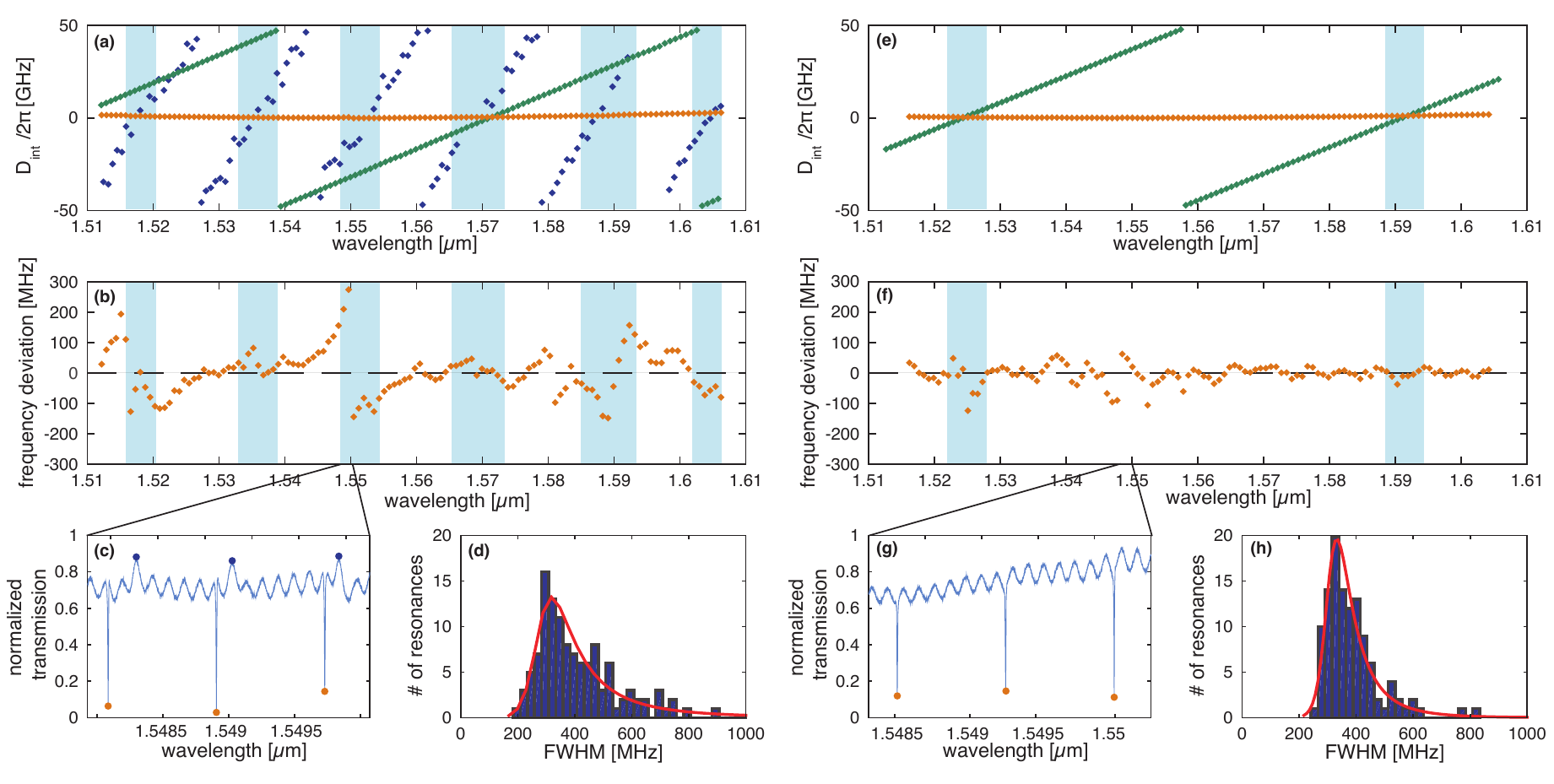}
% Panel A: FSR = 95.967GHz, D_2/2\pi = 1.0958MHz
% Panel B: FSR = 95.995GHz, D_2/2\pi = 0.784MHz
\caption{Characterization of a resonator with constant waveguide width (a-d) comparing to one including the mode filtering section with $w_{\rm t} = 0.45 ~\mu {\rm m}$ (e-h); (a) integrated dispersion $D_{{\rm int}}$ as a function of wavelength of two fundamental modes and one higher order mode families, with ${\tilde D_1 = D_{\rm 1,TM_{00}} \approx 2\pi \times 96 ~{\rm GHz}}$; (b) dispersion deviation of the fundamental TM mode (defined as the deviation of $D_{{\rm int}}$ from a parabolic profile, i.e. $D_{{\rm int}} - \frac{1}{2} D_2 \mu ^2$); (c) microresonator transmission trace around one wavelength; (d) distribution of the resonance linewidth, over the measurement wavelength range, which is fitted by the Burr distribution (red line); (e-h) show the corresponding data as (a-d) for the resonator with filter section.}
\label{dispersionMeasurementData}
\end{figure*}

SiN microresonators were characterized using frequency comb assisted tunable laser spectroscopy\citep{Riemensberger2012dispMeasurement}.
Figure \ref{fig:SetupAndQs}(a) shows the setup that makes use of two beat signals -- the probe (ECDL) with (1) a self referenced and phase stabilized fiber laser frequency comb and with (2) a continuous wave laser -- to calibrate the frequency axis and therefore, precisely measure resonance frequencies and linewidths of the microresonator.
A fiber polarization controller at the microresonator input enables the measurement of mode families of all polarizations.

Resonance frequencies ${\omega _\mu}$ are defined with respect to a central resonance frequency $\omega_0$ as ${\omega_\mu = \omega_0 + D_1 \mu + \frac{1}{2} D_2 \mu ^2 + \cdots}$, where $\mu$ is the relative mode number, ${\frac{D_1}{2\pi}}$ is the FSR, ${D_2 = -\frac{c}{n}{D_1^2}\beta_2}$ is the second order dispersion parameter. A positive-valued $D_2$ implies that the microresonator is in the anomalous GVD regime with ${\beta_2 <0}$. Integrated dispersion ${D_{\rm int}}$ is described as the deviation of the resonance frequencies compared to an equidistant $\tilde D_1$-spaced grid, i.e. ${D_{\rm int} = \omega _\mu - \omega _0 - \tilde D_1 \mu}$.

We compare the characterization of a standard resonator with constant waveguide width (Fig. \ref{dispersionMeasurementData}(a-d)) to one including the mode filtering section with $w_{\rm t} = 0.45 ~\mu {\rm m}$ (tapered resonator, Fig. \ref{dispersionMeasurementData}(e-h)).
Mode family spectra are detected for both resonators (Fig. \ref{dispersionMeasurementData}(a,e)).
We identified two fundamental mode families and one higher order mode for the standard resonator, i.e. ${\rm TE_{00}}$, ${\rm TM_{00}}$ and ${\rm TE_{10}}$, while in the tapered resonator the higher order mode is suppressed.
The transmission trace of the standard resonator, Fig. \ref{dispersionMeasurementData}(c), shows the higher order ${\rm TE_{10}}$ mode that has Fano resonance shape \cite{Ding2014Fano}, together with the fundamental ${\rm TE_{00}}$ mode that has narrower resonance linewidths.
The oscillating background is induced by the Fabry-P\'{e}rot interference between the two chip facet.
The ${\rm TE_{00}}$ mode has the largest FSR according to simulations, which is measured to be $\sim 96 ~{\rm GHz}$.
However, in the transmission trace of the tapered resonator, Fig. \ref{dispersionMeasurementData}(g), resonances of ${\rm TE_{10}}$ mode are not observed. The FSR of the two fundamental modes remain approximately unchanged compared to the standard resonator. On the other hand, the ${D_2}$ parameter is reduced by introducing the mode filtering section, since the tapered waveguide section will contribute a small amount of normal GVD to the overall cavity dispersion. In the shown case, ${\frac{D_2}{2\pi}}$ near 1550 nm is reduced from ${\sim}$1.0 MHz (standard waveguide) to $\sim$0.8 MHz (tapered resonator). Still, the tapered resonator is in the anomalous GVD regime that is necessary for the formation of dissipative Kerr soliton.

Since different modes have different FSRs, they show distinct slopes in the mode spectrum.
Therefore, mode families cross with each other, implying resonances of two mode families are getting close at certain frequencies (crossing points), see Fig. \ref{dispersionMeasurementData}(a,e).
It is at such frequencies that avoided modal crossings appear. This can lead to strong local deviations from the parabolic curvature of the integrated dispersion, see Fig. \ref{dispersionMeasurementData}(b) for the ${\rm TE_{00}}$ mode family of the standard resonator.
In the tapered resonator, avoided modal crossings are much suppressed, see Fig. \ref{dispersionMeasurementData}(f), by filtering out the higher order mode.

The resonance linewidth distribution was also investigated over the whole measurement wavelength range 1.51 -- 1.61 ${\mu {\rm  m}}$, see Fig. \ref{dispersionMeasurementData}(d,h). Both standard and tapered resonators have a similar distribution that can be fitted with a general Burr probability distribution function \cite{burr1942} in order to account for the mean value and the standard deviation of the resonance linewidth.
A careful investigation of the resonance linewidth over different waveguide widths is shown in Fig. \ref{fig:SetupAndQs}(b), with all resonators having the same coupling geometry and being almost critically coupled. This demonstrates that including a modal filtering section in SiN microresonators will not degrade the resonance \emph{Q} factor (for values of $10^6$).

%\section{Dissipative Kerr soliton based frequency comb generation}

\begin{figure}[t]
\includegraphics[width=\linewidth]{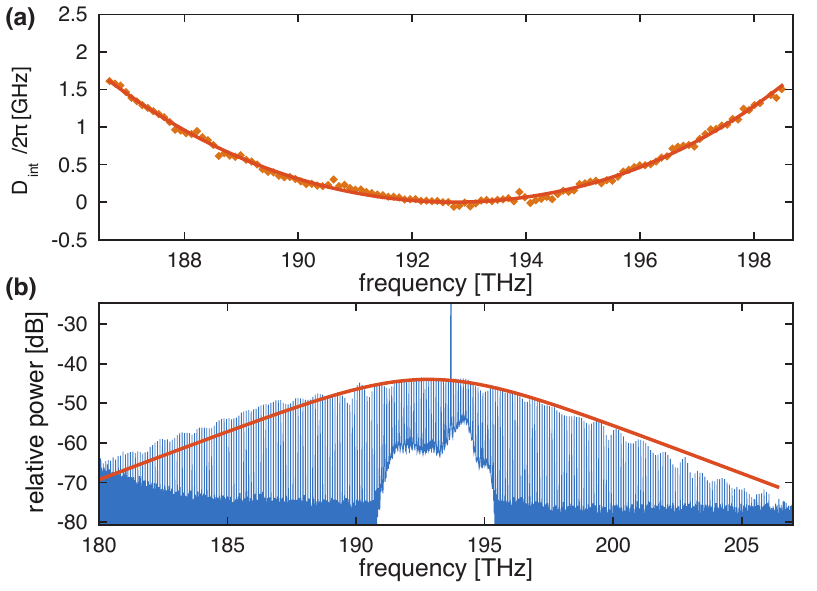}
\caption{(a) Integrated dispersion $\frac{D_{\rm int}}{2\pi}$ of a SiN microresonator with filtering section, with the taper waist ${w_{\rm t} = 0.5 ~\mu {\rm m}}$ yielding a measured anomalous GVD parameter of ${\frac{D_{2}}{2\pi} \approx \mathrm{0.8 ~MHz}}$; (b) frequency comb with single dissipative Kerr soliton formation; the spectrum is fitted with a $\mathrm{sech}^{2}$ function; the 3-dB bandwidth is around 6.4 THz; the soliton comb envelope has an offset of $\sim$0.95 THz from the pump.}
\label{fig:soli}
\end{figure}

We next show the Kerr frequency comb generation based on temporal dissipative Kerr soliton formation in a SiN microresonator with filtering section.
The laser detuning scheme introduced in \cite{Herr2013firstSoliton} was applied, in which the pump laser frequency is swept over a resonance of a fundamental mode and is stopped when the frequency comb generation is in the stable soliton state. One can also apply the \eqt{power kick} used in a previous work on soliton formation in SiN resonators \cite{Brasch2014Cherenkov}.
Figure \ref{fig:soli}(b) shows the generated soliton comb in a 100 GHz microresonator with a taper waist ${w_{\rm t} = 0.5 ~\mu {\rm m}}$.
The corresponding integrated dispersion $\mathrm{D_{\rm int}}$ is shown in Fig. \ref{fig:soli}(a).
The experiment employed 1 W of continuous wave pump light (at 1548 nm) in the waveguide.
The soliton comb has a spectral span of $\sim$25 THz. The 3-dB bandwidth is $\sim$6.4 THz corresponding to the Fourier limited pulse duration of $\sim$48 fs.
The frequency comb spectral envelope is fitted with a $\mathrm{sech^2}$ profile and reveals a slight asymmetry.
The asymmetric spectral envelope is attributed to the third order dispersion ($\frac{D_3}{2\pi} = \mathcal{O}(1)~\mathrm{kHz}$) as well as the self-steepening effects of the microresonator, which induces asymmetry in the parabolic curvature in the dispersion.
Moreover the soliton spectral envelope shows an offset of ${\sim 0.95 ~{\rm THz}}$ from the pump, which is attributed to the Raman induced soliton self-frequency shift as the intracavity soliton is estimated to have an intense peak power ($\mathcal{O}(1)~\mathrm{kW}$) \cite{Karpov2015raman}.
%The experiment employed 1 W of continuous wave pump in the waveguide.

\section{Conclusion}

We have introduced a new resonator layout featuring a single mode filtering section for an integrated SiN platform. This design preserves the high \emph{Q} and the anomalous GVD of the silicon nitride resonator, and effectively suppresses avoided modal crossings caused by the interaction of higher order transverse modes of the waveguide of the micro-ring resonator. This realizes an effectively single-mode micro-ring resonator with anomalous GVD. The new devices show significantly reduced local dispersion deviation due to avoided crossings. It is shown that the design enables reliable generation of temporal dissipative Kerr solitons. The approach is particularly useful for low free-spectral range resonators, or resonators with large number of transverse modes.

\section*{Funding Information}
%\emph{Funding information} ---
This publication was supported by Contract W31P4Q-14-C-0050 from the Defense Advanced Research Projects Agency (DARPA), Defense Sciences Office (DSO).
This work was also supported by the Switzerland National Science Foundation (SNSF). VB acknowledges the support of the European Space Agency (ESA).
\section*{Acknowledgments}
SiN microresonator samples were fabricated in the EPFL nanofabrication facility (CMi).

%\section*{References}

\bigskip

%\bibliography{references}

\end{document}